\documentstyle[aps]{revtex}
\def\narrowtext{}
\def\pacs#1{\centerline{PACS: #1}}

\def\CQG{Class.\ Quant.\ Grav.}
\def\GRG{Gen.\ Rel.\ Grav.}
\def\PRD{Phys.\ Rev.\ {\bf D}}
\def\Lie#1{\hbox{\pounds}\!_{\scriptscriptstyle #1}}
\def\ip#1{i_{\scriptscriptstyle #1}}
\def\div{{\rm div}}

\begin{document}


\title{
\hfill{gr-qc/9404002} \\
 \hfill{\rm 31 March 1994} \\[5mm]
  {\bf THE PATCHWORK DIVERGENCE THEOREM}
  }

\author{Tevian Dray}
\address{Department of Mathematics, Oregon State University,
		Corvallis, OR  97331, USA \\
{\tt tevian{\rm @}math.orst.edu} \\[-1mm]
}

\author{Charles Hellaby}
\address{Department of Applied Mathematics,
		University of Cape Town, Rondebosch 7700, SOUTH AFRICA \\
{\tt cwh{\rm @}maths.uct.ac.za} \\[2.5mm]
}

\maketitle

\widetext


\begin{abstract}
The divergence theorem in its usual form applies only to suitably smooth
vector fields.  For vector fields which are merely piecewise smooth, as is
natural at a boundary between regions with different physical properties, one
must patch together the divergence theorem applied separately in each region.
We give an elegant derivation of the resulting {\it patchwork divergence
theorem} which is independent of the metric signature in either region, and
which is thus valid if the signature changes.
\end{abstract}

\narrowtext

\section{INTRODUCTION}

In previous work \cite{Failure} we discussed the failure, in the presence of
signature change, of the standard conservation laws normally derived from
Einstein's equations.  In the process, we derived a form of the divergence
theorem which is applicable when the signature changes, and related it to the
usual divergence theorem across a boundary surface, i.e.\ for piecewise smooth
vector fields.  We here give a combined treatment of these results which
emphasizes their similarity.

After first establishing our notation in Section II, we discuss the divergence
theorem in the absence of a metric and relate it to the more usual formulation
when a metric is given.  In Section III we apply the divergence theorem to
piecewise smooth vector fields, obtaining the {\it patchwork divergence
theorem}.  In Section IV we show how the patchwork divergence theorem
generalizes standard results on boundary surfaces to our previous results on
signature change.  Finally, in Section V, we discuss some subtle issues
related to the choice of differentiable structure in the presence of signature
change.

\section{THE USUAL DIVERGENCE THEOREM}

The divergence theorem is usually stated in the presence of a (nondegenerate)
metric.  However, there is an alternate formulation which only requires a
volume element, which we now summarize.

A {\it volume element} on an $n$-dimensional oriented manifold $M$ is a
nowhere vanishing $n$-form which is compatible with the orientation on $M$.
In local coordinates $x^i$, an orientation is determined by choosing one of
$\pm dx^1 \wedge ... \wedge dx^n$, and a volume element $\omega$ can then be
obtained by multiplying this by any strictly positive function $\omega_0$.
Assuming a suitable ordering of the coordinates, we thus have
\begin{equation}
  \omega = \omega_0 \: dx^1 \wedge ... \wedge dx^n
\end{equation}
Given a volume element $\omega$ and a vector field $X$ on $M$, we define the
{\it divergence of $X$} by
\begin{equation}
  \div(X) \; \omega := \Lie{X} \omega
\end{equation}
Using the standard expression for the Lie derivative $\Lie{X}\omega$ of
$\omega$ in terms of components with respect to local coordinates $x^i$ (see
e.g.\ \cite{BG}), namely
\begin{equation}
\left( \Lie{X} \omega \right)_{ab...e} =
			X^c \, \partial_c \omega_{ab...e}
			+ \omega_{cb...e} \, \partial_a X^c
			+ \omega_{ac...e} \, \partial_b X^c
			+ ...
			+ \omega_{ab...c} \, \partial_e X^c
\end{equation}
this can be written
\begin{equation}
  \div(X) = {{X^c \partial_c \omega_0}\over{\omega_0}} + \partial_c X^c
\label{Lie}
\end{equation}
Note that if a (nondegenerate)
metric $g$ is given, if $\omega$ is assumed to be the metric volume element
(so $\omega_0=\sqrt{|\det(g)|}$), and if $\nabla$ denotes the Levi-Civita
connection determined by $g$, then our definition agrees with the standard
one, namely
\begin{equation}
  \div(X) = \nabla_a X^a
\end{equation}
This follows since the torsion-free property of $\nabla$ allows us to replace
partial derivatives by covariant ones in (\ref{Lie}), while
metric-compatibility means that $\nabla_a(\det{g})$ vanishes.
We will refer to this as the {\it physical divergence} because of the
important role played by the metric when making physical measurements.

Given an open region $W$ of $M$ bounded by $S$, Stokes' theorem says
that
\begin{equation}
  \oint_S \alpha = \int_W d\alpha
\end{equation}
for any (suitably smooth) ($n-1$)-form $\alpha$, where $S$ must have the
orientation induced by $W$.  Using the identity relating Lie differentiation
to exterior differentiation $d$ and the interior product $\ip{}$, namely (see
e.g.\ \cite{BG})
\begin{equation}
  \Lie{X}\alpha = d(\ip{X}\alpha)+\ip{X}(d\alpha)
\end{equation}
for any differential form $\alpha$, and noting that $d\omega=0$, leads to the
following preliminary form of the divergence theorem
\begin{equation}
  \int_W \div(X) \; \omega = \oint_S \ip{X} \omega
\end{equation}
In order to relate this to the usual divergence theorem, we need to rewrite
the RHS in terms of the induced volume element on $S$.

We will make the customary identification of the tangent and cotangent spaces
of $S$ with the corresponding subspaces of those of $M$, so that in particular
we have
\begin{equation}
  T_p S \subset T_p M
  \qquad {\rm and} \qquad
  T^*_p S \subset T^*_p M \qquad\qquad (p\in S)
\end{equation}
We define a 1-form $m\in T^*_p M$ for $p\in S$ to be {\it normal to $S$} if
\begin{equation}
  m(Y\!)=0 \qquad\qquad (\forall\,Y\in T_p S)
\end{equation}
We note that if $m\ne0$ is normal to $S$ then each $Y\in T_p M$ such that
$m(Y\!)\ne0$ is not tangent to $S$, and hence is either inward or outward
pointing.  We further define $m$ to be {\it outward pointing} if $m(Y\!)>0$
whenever $Y$ is outward pointing, and {\it inward pointing} if $m(Y\!)<0$ for
all
such $Y$.  These definitions extend directly from vectors and 1-forms at a
point to the corresponding tensor fields.  To give a simple and common
example, if $W=\{f<0\}$ and $S=\{f=0\}$, then $df$ is normal to $S$ and
outward pointing.  In fact, the 1-forms which are normal to $S$ and outward
pointing are precisely the positive multiples of $df$.

Given a 1-form $m$ which is normal to $S$, we can define an ($n-1$)-form
$\sigma$ on $S$ via
\begin{equation}
  m \wedge \sigma := \omega
\label{Induced}
\end{equation}
The normality of $m$ ensures that there is a unique $\sigma$ on $S$ satisfying
this equation.  Since $\omega$ is compatible with the orientation on $M$,
$\sigma$ is compatible with the induced orientation on $S$ precisely when $m$
is outward pointing.  In this case, $\sigma$ is the {\it volume element on $S$
induced by $m$ and $\omega$}, or more simply the {\it induced volume element
on $S$}.  The interior product is a derivation, so that in particular
\begin{equation}
  \ip{X} (m \wedge \sigma) = (\ip{X}m) \wedge \sigma - m \wedge (\ip{X}\sigma)
\end{equation}
Note that $\ip{X}m=m(X)$ and that the pullback of $m$ to $S$ is zero.

Putting this all together, we finally obtain the {\it divergence theorem} in
the form
\begin{equation}
  \int_W \div(X) \; \omega = \oint_S m(X) \; \sigma
\end{equation}
where $m$ is (any) outward pointing normal to $S$ and $\sigma$ is the induced
volume element defined above.

In the presence of a metric and assuming that $S$ is not null, letting $m_a$
denote the (components of the) outward unit normal 1-form to $S$ leads to the
{\it physical divergence theorem}, namely
\begin{equation}
  \int_W \nabla_a X^a \; \omega = \oint_S X^a m_a \; \sigma
\end{equation}
where $\sigma$ is the metric volume element on $S$.  It is customary to write
this as
\begin{equation}
  \int_W \nabla_a X^a \; d^n\!W = \oint_S m^a X_a \; d^{n\!-\!1}\!S
\end{equation}
where $d^n\!W=\omega$ and $d^{n\!-\!1}\!S=\sigma$ denote the metric volume
elements on $W$ and $S$, respectively, and $m^a$ denotes the (components of
the) vector field which is the metric dual of $m_a$; we will refer to this
formulation as the {\it physical divergence theorem}.  It is important to note
that while $m^a$ is indeed a unit vector normal to $S$, it is outward pointing
only when it ($m^a$) is spacelike, and is instead inward pointing where it is
timelike; $m_a$ is of course always outward pointing.

\section{THE PATCHWORK DIVERGENCE THEOREM}

Consider now a boundary surface $\Sigma$, which divides $W$ and $S$ into two
parts $W^+$ \& $W^-$ and $S^+$ \& $S^-$, and let $S^0 = \Sigma \cap W$ be the
enclosed region of $\Sigma$.
\footnote{We emphasize that $\Sigma$ is to be viewed as a hypersurface in a
given manifold $M$, so that there are no complications in the manifold
structure at or near $\Sigma$; this issue is discussed further in Section
\ref{Discussion}.}
Suppose an outward-pointing 1-form $m$ is given on $S$, and further suppose
that outward-pointing 1-forms $m^\pm$ are given on the separate boundaries
$\partial W^\pm = S^\pm\cup S^0$ of the two regions $W^\pm$ which agree with
$m$ on $S$ and which are equal but opposite on $\Sigma$, i.e.\
\begin{eqnarray}
  m^\pm|_{S^\pm} & = & m \\
  m^-|_\Sigma & = & - m^+|_\Sigma =: l
\end{eqnarray}
Let $\omega$ and $\sigma$ denote as usual the metric volume elements on $M$
and $S$, and let $\sigma^\pm$ denote the induced volume elements on the
boundaries $\partial W^\pm$, so that
\begin{eqnarray}
  \sigma^\pm|_S & = & \sigma \\
  \sigma^-|_\Sigma & = & - \sigma^+|_\Sigma =: \sigma^0
\end{eqnarray}
We give $S^0$ the orientation induced by $l$ and $\omega$, namely $\sigma$,
which is the orientation it inherits as part of the boundary of $W^-$ (and not
$W^+$).

Now consider a vector field $X$ which is piecewise smooth, so that the usual
divergence theorem can be applied in each region.  Adding the two resulting
equations gives
\begin{eqnarray}
         \int_{W^+} \div(X_+) \; \omega
   & + & \int_{W^-} \div(X_-) \; \omega \nonumber\\
   & = & \oint_{\partial W^+} m^+(X) \; \sigma^+
       + \oint_{\partial W^-} m^-(X) \; \sigma^- \nonumber\\
   & = & \int_{S^+} m^+(X_+) \; \sigma^+
       - \int_{S^0} m^+(X_+) \; \sigma^+
       + \int_{S^-} m^-(X_-) \; \sigma^-
       + \int_{S^0} m^-(X_-) \; \sigma^- \nonumber\\
   & = & \int_{S^+} m(X) \; \sigma^+
       + \int_{S^-} m(X) \; \sigma^-
       - \int_{S^0} l([X]) \; \sigma^0
\end{eqnarray}
where the minus sign in the third line is due to the difference in orientation
of $S^0$ and $\partial W^+$ and where
\begin{equation}
   [Q] := \lim_{\to\Sigma^+} Q^+ - \lim_{\to\Sigma^-} Q^-
\end{equation}
denotes the discontinuity in $Q$ across $\Sigma$.  We can rewrite this as
\begin{equation}
    \int_{W} \div(X) \; \omega
  = \oint_S m(X) \; \sigma
  - \int_{S^0} l([X]) \; \sigma^0
\label{Patch}
\end{equation}
and this is the {\it patchwork divergence theorem}.  In components we obtain
\begin{equation}
    \int_{W} \nabla_a X^a \; \omega
  = \oint_S X^a m_a \; \sigma
  - \int_{S^0} \left[ X^a \right] l_a \; \sigma^0
\label{pdt}
\end{equation}
In the presence of a metric, it is customary to assume that $m$ and $l$ are
unit; note that the convention adopted here is that $l$ points from $W^-$ to
$W^+$.  If we introduce the (unit) vector field $n^a$ which is normal to
$\Sigma$ and which points from $W^-$ to $W^+$, we finally obtain
\begin{equation}
    \int_{W} \nabla_a X^a \; d^n\!W
  = \oint_S m^a X_a \; d^{n\!-\!1}\!S
  - \epsilon\int_{S^0} n^a \left[ X_a \right] \; d^{n\!-\!1}\!\Sigma
\label{PDT}
\end{equation}
where $d^{n\!-\!1}\Sigma=\sigma^0$ denotes the metric volume element on
$\Sigma$ and where there is an important sign difference depending on whether
$n^a$ is spacelike ($\epsilon=1$) or timelike ($\epsilon=-1$); $l_a n^a = 1$
in both cases.

\section{APPLICATIONS}

\subsection{Boundary Surface}

A boundary surface in general relativity can be represented as a hypersurface
in
a Lorentzian manifold across which the matter model changes.  A spacelike
hypersurface corresponds to a change in the matter model at a particular time,
and a timelike hypersurface corresponds to a change at a particular place,
while a null hypersurface corresponds to a gravitational shock wave.  We thus
consider a manifold $M$ with a Lorentzian metric $g$ and a given non-null
hypersurface $\Sigma$ which divides $M$ into 2 regions $M^\pm$.  Let $n$
denote the unit normal vector to $\Sigma$ which points from $M^-$ to $M^+$,
and let $\epsilon=g(n,n)=\pm1$.  The extrinsic curvature $K$ of $\Sigma$ can
be defined by
\begin{equation}
  2K:=\hbox{\pounds}_n g
\end{equation}
so that in components
\begin{equation}
  2 K_{ab} = n^c \partial_c g_{ab}
             + g_{mb} \partial_a n^c + g_{am} \partial_b n^c
\end{equation}
We shall assume here that $g$ is $C^1$ across $M$, so that in particular the
{\it Darmois junction conditions}
\begin{equation}
  [ h ] = 0 = [ K ]
\label{Darmois}
\end{equation}
on the induced metric $h=g+\epsilon n\otimes n$ and extrinsic curvature $K$ of
$\Sigma$ are satisfied.
\footnote{We reiterate that we are assuming that $\Sigma$ is a hypersurface in
a given manifold, so that local coordinates exist which span $\Sigma$.
Equation (\ref{Darmois}) is thus to be interpreted as applying to
$n$-dimensional tensors, obtained by projection into $\Sigma$.  In practice,
however, it is often convenient to consider the pullback of Equation
(\ref{Darmois}) to $\Sigma$, which contains the same information but which
only requires local coordinates {\it within} $\Sigma$.}

In the non-null case, Israel \cite{Israel} used Einstein's equations and the
Gauss-Codazzi relations between the curvature of $M$ and that of $\Sigma$ to
relate the stress-energy tensor of the matter to the intrinsic and extrinsic
curvatures of $\Sigma$.  Clarke and Dray \cite{Null} generalized some of
Israel's results to the null case.  Corrected versions of some of Israel's
result appear in \cite{Failure}, including
\begin{equation}
  \rho := G_{ab} \: n^a n^b = \frac{1}{2}
    \left( (K^c{}_c)^2 - K_{ab} K^{ab} - \epsilon \: {\cal R} \right)
\end{equation}
where ${\cal R}$ denotes the scalar curvature of the intrinsic metric $h$ on
$\Sigma$.  In the case where $n^a$ is timelike, $\rho$ can be interpreted as
the energy density.  If the Darmois junction conditions are satisfied, the RHS
of this equation is continuous at $\Sigma$, and therefore so is the energy
density, yielding
\begin{equation}
  [ \rho ] = \left[ G_{ab} \: n^a n^b \right] = 0
\label{energy}
\end{equation}

If we now apply the patchwork divergence theorem (\ref{PDT}) to
\begin{equation}
  X_a := G_{ab} n^b
\end{equation}
then the term containing $l([X])$ vanishes, and we are left with the usual
statement of the divergence theorem as though no boundary surface were
present.  If we instead consider
\begin{equation}
  X_a := G_a{}^b l_b = l_b G^b{}_a
\end{equation}
we obtain the same result.  It is important to note that these results hold
even though $X$ may well not be $C^1$!

The results of Dray and Padmanabhan on piecewise Killing vectors
\cite{Killing} can be viewed as application of the patchwork divergence theorem
in this setting.

\subsection{Signature Change}

Consider now a situation similar to the above, but where the metric is
Lorentzian only in $M^-$, and Riemannian in $M^+$, so that the metric is now
either discontinuous or zero at $\Sigma$.  Such signature-changing models were
introduced by Dray {\it et al.}\ \cite{PaperI,PaperII,PaperIII}; similar
models have since been used in a cosmological setting
\cite{EllisA,EllisB,Hayward,Dereli,Schwarz}.

We can impose the Darmois junction conditions by means of 1-sided limits to
$\Sigma$.  In order to apply the patchwork divergence theorem we need a volume
element on $\Sigma$, and we cannot now use the metric volume element there.
One possibility is to work in normal coordinates and note that both sides
induce the same volume element; as suggested in \cite{Failure}.  We discuss
this issue further in the next section, and here assume that a suitable choice
has been made.

On either manifold-with-boundary $M^\pm$, let $n^a$ denote the unit normal
vector as described just before and after (\ref{PDT}).
\footnote{We will assume throughout this section that these normal vectors make
sense, and can be obtained as 1-sided limits from $M^\pm$.  This important
issue will be discussed in Section \ref{Discussion}, where one way of
achieving this is described.  As discussed there, what really matters is
whether the scalars in the last term in (\ref{PDT}) have limits, not whether
the normal vectors $n^a$ do.}
Even though $\epsilon$ now changes sign between the two regions, $\epsilon
n^a$ is continuous across $\Sigma$, so that the RHS of (\ref{PDT}) remains
well-defined.  However, Israel's results such as (\ref{energy}) no longer
hold, and one obtains instead \cite{Failure}
\begin{eqnarray}
\label{MatchA}
  \left[ G_{ab} \: n^a n^b \right] & = & - {\cal R} \\
  \left[ G^a{}_b \: n^b l_a \right] & = & (K^c{}_c)^2 - K_{ab} K^{ab}
\label{MatchB}
\end{eqnarray}

The patchwork divergence theorem now contributes a surface term at $\Sigma$,
and for $X_a=G_{ab}n^b$ as above we now obtain
\footnote{In deriving this result it is slightly easier to use (\ref{MatchB})
in (\ref{pdt}) rather than using (\ref{PDT}) directly.}
\begin{equation}
    \int_{W} \nabla_a X^a \; d^n\!W
  = \oint_S m^a X_a \; d^{n\!-\!1}\!S
  - \int_{S^0} \; \left( (K^c{}_c)^2 - K_{ab} K^{ab} \right)
d^{n\!-\!1}\!\Sigma
\label{conservation1}
\end{equation}
If, however, we instead set $X_a=G_a{}^bl_b$, we now obtain a different
conservation law, namely
\begin{equation}
    \int_{W} \nabla_a X^a \; d^n\!W
  = \oint_S m^a X_a \; d^{n\!-\!1}\!S
  + \int_{S^0} \; {\cal R} \; d^{n\!-\!1}\!\Sigma
\label{conservation2}
\end{equation}
These are two of the main results of \cite{Failure}.

\section{DISCUSSION}
\label{Discussion}

In the {\it normal coordinate} approach, $M^\pm$ are viewed as disjoint
manifolds-with-boundary, with metrics of different signatures, which are being
identified via an isometry of their boundaries $\Sigma^\pm$.  To make the
result a manifold, a differentiable structure must be specified at the
identified boundary $\Sigma$.  This can naturally be done by requiring that
normal coordinates, defined separately on either side of $\Sigma$, be
admissible coordinates.  In these coordinates, a signature-changing metric
will take the form
\begin{equation}
      ds^2 = \epsilon \, d\tau^2 + h_{ij} \, dx^i dx^j
\end{equation}
where $\tau$ denotes proper time/proper distance away from $\Sigma$, and
$\{x^i: i=1,...,n-1\}$ are local coordinates on $\Sigma$.  Although this
metric is discontinuous at $\Sigma$, the metric volume elements
\begin{equation}
 d^n\!V := \sqrt{|\det(g)|} \: d\tau \wedge dx^1 \wedge ... \wedge dx^{n\!-\!1}
\end{equation}
obtained separately on $\Sigma$ using 1-sided limits from $M^\pm$ are
identical, so that
\begin{equation}
 d^{n\!-\!1}\!\Sigma := \sqrt{|\det(g)|} \: dx^1 \wedge ... \wedge dx^{n\!-\!1}
\end{equation}
can be taken to be the natural volume element on $\Sigma$.  Furthermore, in
this approach the normal vectors $n^a$ can clearly be obtained as 1-sided
limits to $\Sigma$.  This leads to the results of the previous section.

On the other hand, one can take the {\it continuous metric} approach, in which
one starts with a manifold $M$ on which a continuous, covariant tensor $g$ of
rank two is given which is assumed to be a metric on (the interior of)
$M^\pm$.  The only way for the signature of $g$ to change is for it to be
degenerate at $\Sigma$; we nonetheless refer to $g$ as a {\it
signature-changing metric}.  Such a metric can be put in the form
\begin{equation}
      ds^2 = N \, dt^2 + h_{ij} \, dx^i dx^j
\end{equation}
where the function $N$ is zero (only) on $\Sigma$.  The limit of the metric
volume element on $M$ to $\Sigma$ is now zero, so that it is not clear how to
interpret (\ref{Induced}) for the induced volume element.  Nonetheless,
$\Sigma$ does still have a metric volume element due to the induced
(nondegenerate) metric $h$, and this volume element agrees with
$d^{n\!-\!1}\!\Sigma$ as defined using the normal coordinate approach.

In the continuous metric approach, however, there is no unit normal vector
which can play the role of $m$ in the divergence theorem.  If we pick
\begin{equation}
  m = \sqrt{|N|} \, dt
\end{equation}
so that (\ref{Induced}) is satisfied away from $\Sigma$, then $m=0$ on
$\Sigma$, and the RHS of the divergence theorem is identically zero (provided
$X$ is suitably smooth; see below).  Contrast this with the normal coordinate
approach, in which $m=d\tau$ is a basis 1-form and hence nonzero!  In order to
preserve a divergence theorem in this approach, we are thus led to pick an
arbitrary, nonzero $m$, and then to define the volume element on $M$ via
(\ref{Induced}) rather than using the metric volume element.  The patchwork
divergence theorem in the form (\ref{Patch}) now holds, and can be used to
derive results analogous to those in the previous section, although the volume
integral over $W$ is nonstandard.

Which of these approaches is relevant depends of course on what problem is
being discussed.  In particular, the very requirement that the tensors in the
integrands of the divergence theorem be suitably smooth is different in the
two approaches.  For instance, {\bf if} $X^a$ is assumed to be smooth at
$\Sigma$ in the continuous metric approach, {\bf then} $m_a X^a=0$ in that
approach.  If, on the other hand, $X^a$ is smooth in the normal coordinate
approach, $m_a X^a$ can have a nonzero limit at $\Sigma$.
\footnote{This problem is in addition to the necessity of specifying which
index structure for a given tensor is fundamental, as shown e.g.\ by the
difference between Equations (\ref{conservation1}) and (\ref{conservation2}).}
In a given physical situation with given smoothness of the physical fields,
the same physical results will be obtained in either approach (provided
suitable limits are taken).  However, we feel that, due to the fundamental
role played by unit normal derivatives in initial-value problems, the normal
coordinate approach is more likely to automatically incorporate appropriate
smoothness conditions at $\Sigma$.

Clarke \& Dray \cite{Null} showed that the continuity of the induced metric
$h$ at the identified boundary of two manifolds-with-boundary leads to a
unique differentiable structure for the resulting manifold.  While their
derivation assumed constant signature, this was only used in showing that the
resulting $n$-dimensional metric is continuous.  Thus, their result naturally
extends to the case of signature change, where it results in the
differentiable structure defined by the normal coordinate approach (with its
discontinuous metric).

Their result can also be applied to the continuous metric approach as follows.
Suppose that one is given two manifolds-with-boundary with (suitably
controlled) degenerate metrics on the boundaries.  Switching to the
differentiable structures on each side induced by (1-sided) normal coordinates
leads to a unique differentiable structure via the generalized result of
Clarke \& Dray.  One can now reverse at least one of the changes in
differentiable structure, obtaining a manifold structure on the glued-together
manifold which agrees with at least one of the manifolds-with-boundary.  If
(and only if) the degeneracies on the two original boundaries are compatible
in an appropriate sense (based essentially on the behavior of the metric when
expanded as a power series in proper time/distance from $\Sigma$), a unique
manifold structure compatible with both original manifolds-with-boundary is
thus obtained.  In particular, one can always construct several inequivalent
manifold structures with continuous (degenerate) metrics, but at most one of
these will agree with both of the original manifold structures on the two
sides.

This issue can be largely ignored by adopting the {\it invariant approach}
used in \cite{Failure}, which is closely related to the elegant approach used
by Carfora and Ellis \cite{Carfora}.  In this approach, one starts by
identifying the disjoint manifolds-with-boundary $M^\pm$ as in the normal
coordinate approach.  With no further assumptions about the manifold structure
at $\Sigma$, the 1-sided unit normal vectors still make sense, so that the
(pullback of the) Darmois junction conditions can still be imposed.

In this approach, and further assuming that the 1-sided limits of $n^a X_a$
exist, one obtains the patchwork divergence theorem in the form
\begin{equation}
    \int_{W} \nabla_a X^a \; d^n\!W
  = \oint_S m^a X_a \; d^{n\!-\!1}\!S
  - \int_{S^0} \left[ \epsilon n^a X_a \right] \; d^{n\!-\!1}\!\Sigma
\end{equation}
While in practice this is equivalent to (\ref{PDT}), this form of the theorem
emphasizes that it is the limits of physical quantities to $\Sigma$ which
determine the physics, not the choice (or even existence) of a manifold
structure there.

The related issue of defining hypersurface distributions in the presence of
signature change will be discussed elsewhere \cite{PaperIV}.

\section*{ACKNOWLEDGEMENTS}

TD was partially funded by NSF grant PHY 92-08494.
CH would like to thank the FRD for a research grant.

\end{document}